\documentclass[pre,twocolumn,floatfix,showpacs,superscriptaddress,aps]{revtex4}

\usepackage{amsmath,graphicx}
\usepackage{color}

\begin{document}

\title{Experimental confirmation of chaotic phase synchronization in coupled time-delayed electronic circuits}

\author{D.~V.~Senthilkumar}
\affiliation{ Centre for Dynamics of Complex Systems, University of Potsdam, 14469 Potsdam Germany}%
\affiliation{Potsdam Institute for Climate Impact Research, 14473 Potsdam Germany}%
\author{K. Srinivasan}%
\affiliation{Centre for Nonlinear Dynamics, Bharathidasan University, Tiruchirapalli - 620 024, India}%
\author{K. Murali}%
\affiliation{Department of Physics, Anna University, Chennai, India}
\author{M.~Lakshmanan}%
\affiliation{Centre for Nonlinear Dynamics, Bharathidasan University, Tiruchirapalli - 620 024, India}%
\author{J.~Kurths}%
\affiliation{Potsdam  Institute for Climate Impact Research, 14473 Potsdam Germany}%
\affiliation{Institute for Physics, Humboldt University, 12489 Berlin, Germany}%


\begin{abstract}
We report the first experimental demonstration of chaotic phase synchronization (CPS) 
in unidirectionally coupled time-delay systems using electronic circuits. We have also
implemented experimentally an efficient methodology
for characterizing CPS, namely the localized sets. Snapshots of the evolution of coupled systems and the
sets as observed from the oscilloscope confirming CPS are shown experimentally. 
Numerical results from different approaches, namely phase differences, 
localized sets, changes in the largest Lyapunov exponents and the correlation of 
probability of recurrence ($C_{CPR}$), corroborate the experimental observations.
\end{abstract}

\maketitle

%\section{Introduction}
Chaotic phase synchronization (CPS) refers to the coincidence of characteristic 
time scales of interacting chaotic dynamical systems, while their amplitudes remain 
chaotic and often uncorrelated~\cite{aspmgr2001,sbjk2002}. CPS plays a crucial role 
in understanding a large class of weakly interacting nonlinear dynamical systems and has been
demonstrated both theoretically and experimentally in a wide variety of natural 
systems~\cite{scannell1999,csmgr1998,bbah1999,grenfell2001,ercmt2003,kvvvni2001,dmav2000,ptmgr1998,dmjk2005,rvst2004}. 
Despite our substantial understanding of the phenomenon of CPS and its 
potential applications
in low-dimensional systems, only a very few studies on it have been reported
in time-delayed systems, which  are essentially 
infinite-dimensional in nature~\cite{dvs2006,rsdvs2010}. 
Due to the highly non-phase-coherent chaotic/hyperchaotic attractors
with complex topological properties exhibited
by these systems in general, it is often impossible to estimate the phase explicitly
and to identify CPS. 

Recently, we have introduced a nonlinear transformation to recast the original
non-phase-coherent attractors into smeared limit-cycle attractors to enable to
estimate the phase explicitly and to identify CPS in time-delay model systems
for the first time in the literature~\cite{dvs2006}.
In this paper, we report the first experimental demonstration of CPS in coupled
time-delay systems using electronic circuits. We have also experimentally implemented
the methodology of localized sets~\cite{tpmsb2007} and show that this is a crucial and a general 
framework for characterizing CPS even in non-phase-coherent attractors of time-delay
systems~\cite{dvs2006,rsdvs2010}. Our results will open up the possibility of experimental
realization of CPS in other physical systems  with delay and  to
their potential applications.

\begin{figure}[!ht]
\centering\includegraphics[width=1.0\linewidth,clip]{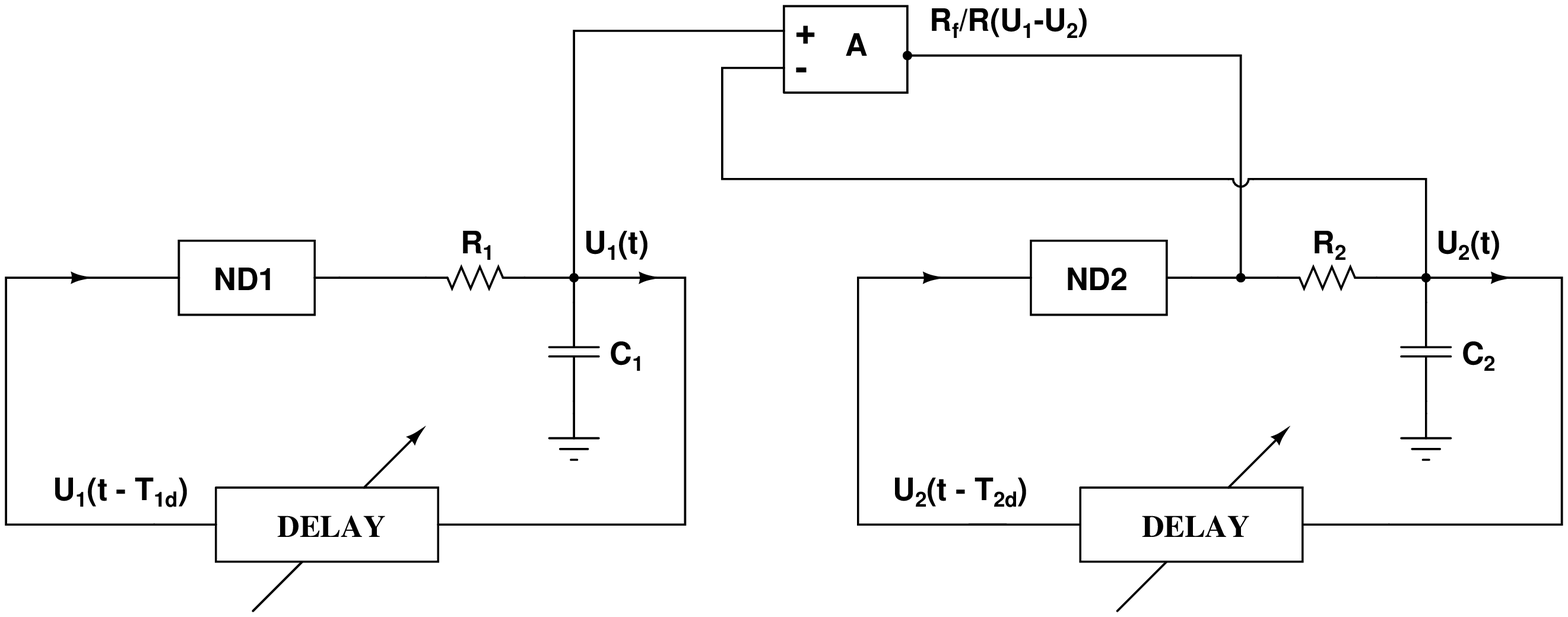}
\caption{Schematic diagram of the unidirectionally coupled time-delay 
analog circuits with threshold nonlinearity.  ND1 and ND2 are nonlinear 
device units; delay unit consists of $10$ pairs of capacitors of $470 nF$ and 
inductors of $12 mH$. $R_1=R_2=1.86 k\Omega$; 
$C_1=C_2=100 nF$ and $A$ is the op-amp difference amplifier.
} \label{blockdiag_coup}
\end{figure}
In particular, we will demonstrate the existence of CPS in unidirectionally coupled
time-delay electronic circuits with threshold nonlinearity in both chaotic
and hyperchaotic regimes experimentally (Note that bidirectional coupling can also
work equally well). In addition to
the snapshots of time series of both systems as seen from the oscilloscope, 
we have used the framework of localized sets~\cite{tpmsb2007} to characterize 
the existence of CPS in the above systems both experimentally and numerically. 
To investigate localized sets, we have considered the `event'
as maxima of the flow of the drive system and recorded the response system to obtain the `sets', 
whenever a maximum occurs in the drive system and vice versa.  The sets are then
superimposed on the drive (response) attractor, which 
get localized on it during CPS but
spread over the entire attractor when the 
systems evolve independently.  Further, we have also confirmed the existence of CPS
numerically using the localized sets, the largest Lyapunov exponents of the coupled time-delay systems
and also with another independent approach based on recurrence analysis, namely
the correlation of probability of recurrence ($C_{CPR}$)~\cite{nmmcr2007}.
\begin{figure}
\centering
\includegraphics[width=0.8\columnwidth]{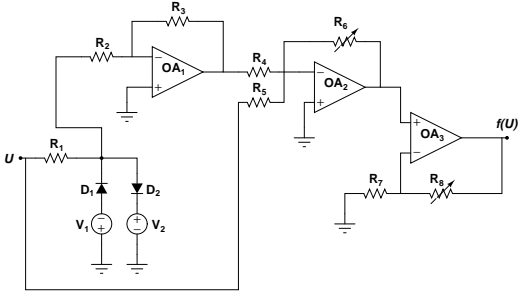}
\caption{\label{fig2b} 
Nonlinear device unit (ND) of Fig.1: Actual circuit implementation of the nonlinear activation function consisting of diodes $D_1=D_2=1N4148$, resistor $R_1=R_7=1 k\Omega, R_2=R_3=10 k\Omega, R_4=2 k\Omega, 
R_5=3 k\Omega, R_6=10.4 k\Omega$ and $R_8=5 k\Omega$ and threshold control voltages $V_1=V_2=0.7V$ along with different amplifying stages ($OA_1=OA_2=OA_3=uA741$).}
%Nonlinear device unit (ND): Circuit implementation of the nonlinear activation function
%with amplifying stages ($OA_2, OA_3$).}
\end{figure}
\begin{figure}[!ht]
%\centering\includegraphics[width=0.8\linewidth]{fig1}
\centering\includegraphics[width=0.7\linewidth,clip]{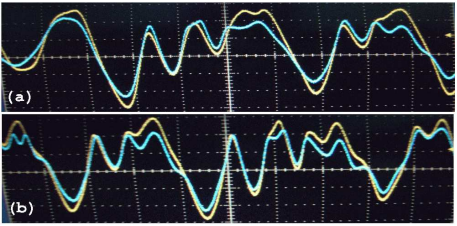}
\caption{(color online) Snapshots of the time evolution of both coupled systems
indicating the existence of CPS in (a) chaotic regime and (b) hyperchaotic regime.
 $x$-axis: time ($1$ unit= $1.0ms$), $y$-axis: voltage ($1$ unit= $1.0V$).} 
\label{time_exp_phase1}
\end{figure}

The coupled electronic circuit investigated here is shown in Fig.~\ref{blockdiag_coup}
as a block diagram. The individual time-delay units have a ring structure and
comprise of a diode based nonlinear device unit (ND) (Fig.~\ref{fig2b}), a variable time-delay 
unit (DELAY) along with an integrator ($R_0C_0$) unit. 

\begin{figure}[!ht]
\centering\includegraphics[width=0.78\linewidth,clip]{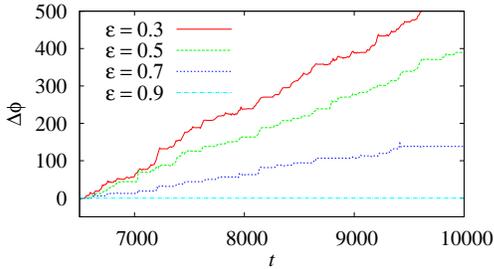}
\caption{(color online) Phase difference for different values of 
$\varepsilon = 0.3, 0.5, 0.7$ and $0.9$ and $\tau=1.33$.} \label{phase_dif}
\end{figure}
The dynamics of the individual circuit in Fig.~\ref{blockdiag_coup} 
is represented by the delay differential equation
$R_0C_0\frac{dU(t)}{dt} = -U(t) + F[k_f U(t-T_d)]$,
where $U(t)$ is the voltage across the capacitor $C_0$, $U(t-T_d)$ is the voltage 
across the delay unit (DELAY), $T_d=n\sqrt{LC}$ is the delay time, $n$ is the number of $LC$ units
and $F[k_f U(t-T_d)]$ is the 
static characteristic of the ND shown in Fig.~\ref{fig2b}.
The block diagram (Fig.~\ref{blockdiag_coup}) also contains a
differential amplifier circuit (A) with a gain $\varepsilon=R_f/R$ used to find the difference
between the two voltage signals $U_1$ and $U_2$.  By changing the feedback 
resistance $(R_f)$, the coupling strength $\varepsilon$ can be varied. 

\begin{figure}[!ht]
%\centering\includegraphics[width=0.8\linewidth]{fig1}
\centering\includegraphics[width=1.0\linewidth,clip]{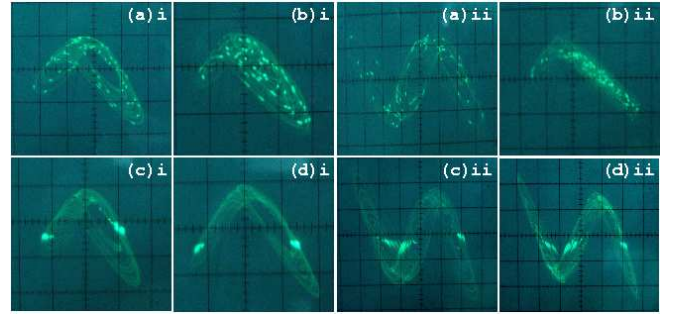}
\caption{(color online) Experimental characterization of CPS 
using the framework of localized sets in the chaotic regime 
(i) for $\tau=1.33$ and in the hyperchaotic regime (ii)  for $\tau=6.0$. 
Sets in the drive and the response systems are distributed in 
(a) and (b) for $\varepsilon=0.3$ indicating the asynchronous state and
localized in (c) and (d) for $\varepsilon=0.9$ indicating CPS, respectively.
$x$-axis : voltage $U(t)$ ($1$ unit= $0.5V$), $y$-axis: voltage $U(t-T_d)$ ($1$ unit= $2.0V$). }
%(a) and (b) The distributed sets in the drive and the response
%systems, respectively, indicate the asynchronous state for $\varepsilon=0.3$. 
%(c) and (d) The localized sets in the drive and the response
%systems, respectively, indicating the existence of CPS for $\varepsilon=0.9$.} 
\label{1b_exp_phas1}
\end{figure}

The normalized evolution equation corresponding
to the coupled time-delay electronic circuits (Fig.~\ref{blockdiag_coup}) 
is represented as~\cite{srini2010,ksdvs2010}
\begin{align}\label{dd_eq}
\dot x = &\, -x(t) + b_1f\left[x(t-\tau)\right], \notag \\
\dot y = &\,  -y(t) + b_2f\left[y(t-\tau)\right]+\varepsilon(x(t)-y(t)),
\end{align}
where $x(t)=y(t)=\frac{U(t)}{U_s}$, $\hat{t}=\frac{t}{R_0C_0}$, $\tau=\frac{T_d}{R_0C_0}$,
and $b=k_f=1+(\frac{R_8}{R_7})$ are dimensionless circuit variables and parameters. 
The function $f(x(t-\tau))=F(U(t-T_d))$ is taken to be a symmetric piecewise linear function 
defined by~\cite{srini2010,ksdvs2010} 

\begin{subequations} 
\begin{equation}
f(x)=Af^* - Bx.
\end{equation}
Here
\begin{eqnarray}
f^*=
\left\{
\begin{array}{cc}
-x^* & \;\;\;\; \;\;  x < -x^*, \\
x &\;\;\;\;\;  -x^* \le x \le x^*, \\
x^*&\;\;\;\; \;\;  x>x^*, 
\end{array} \right.
\end{eqnarray}
\label{non_eq}
\end{subequations}
where $x^*$ is a controllable threshold value
and can be altered by adjusting the values of voltages $V_1$ and $V_2$.
$A=(R_6/R_4)$ and $B=(R_6/R_5)$ are positive parameters.  
The estimated normalized values turn out to be $x^*=0.7$, $A=5.2$, $B=3.5$,  
$b_1=1.2$ and $b_2=1.1$ in accordance with the values of the circuit elements. 
The parameter mismatch $b_1\ne b_2$ contributes to the non-identical nature of 
the coupled time-delay systems. In the following, we will demonstrate 
the existence of CPS as a function
of the coupling strength $\varepsilon$ in both chaotic and hyperchaotic regimes for
suitable values of the delay time $\tau$.
\begin{figure}[!ht]
%\centering\includegraphics[width=1.0\linewidth]{fig1}
\centering\includegraphics[width=1.0\linewidth,clip]{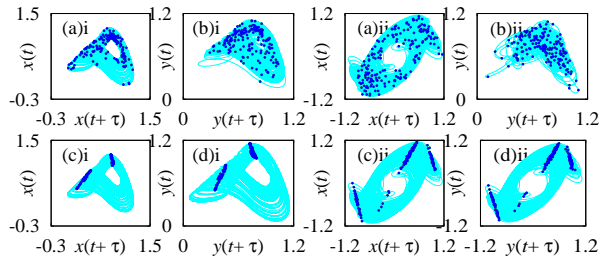}
\caption{(color online) Numerical confirmation of CPS using the framework of
localized sets in the chaotic regime (i) for $\tau=1.33$ and in the hyperchaotic
regime (ii)  for $\tau=6.0$. 
Sets in the drive and the response
systems are distributed in (a) and (b) for $\varepsilon=0.3$ 
indicating the asynchronous state and
localized in (c) and (d) for $\varepsilon=0.9$ indicating CPS, respectively.}
\label{chaos_ls}
\end{figure}

The snapshots of the time series of both drive and response systems as seen from 
the oscilloscope are shown in Fig.~\ref{time_exp_phase1}(a) in the chaotic regime for
the delay time $\tau=1.33$ and the coupling strength $\varepsilon=0.9$,
indicating the evolution of both systems in-phase with each other. Similarly, 
the snapshots of the time series evolving in-phase with each other in the hyperchaotic regime for
the delay time $\tau=6.0$  are shown in Fig.~\ref{time_exp_phase1}(b) for 
$\varepsilon=0.7$. The phase differences calculated  numerically from  the evolution
equations, Eq.~(\ref{dd_eq}), using the Poincar\'e section
technique~\cite{aspmgr2001,sbjk2002} for different values of $\varepsilon$ are
illustrated in Fig.~\ref{phase_dif}, indicating the existence of CPS for 
$\varepsilon=0.9$ with $\tau=1.33$.
The existence of CPS is further characterized both 
experimentally and numerically by using the framework of localized sets~\cite{tpmsb2007}.

\begin{figure}[!ht]
%\centering\includegraphics[width=0.8\linewidth]{fig1}
\centering\includegraphics[width=0.75\linewidth,clip]{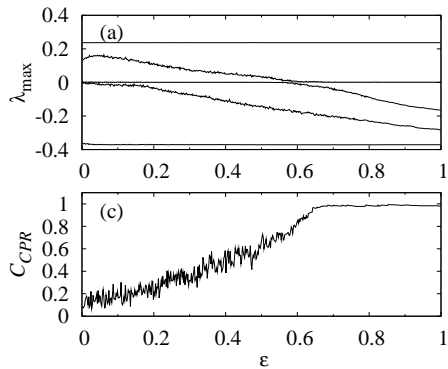}
\caption{Confirmation of CPS in the chaotic regime using  
(a) Four largest Lyapunov exponents and 
(b) Correlation of probability of
recurrence ($C_{CPR}$).} \label{chaos_ps_char}
\end{figure}
The  sets obtained by sampling the time series of one of the systems whenever a maximum
occurs in the other one are plotted along with the chaotic attractor of the same system for the
delay time $\tau=1.33$ both experimentally and numerically in 
Figs.~\ref{1b_exp_phas1} and \ref{chaos_ls}, respectively.  The sets distributed
over the entire attractor of both the drive [Figs.~\ref{1b_exp_phas1}(a)i and 
\ref{chaos_ls}(a)i] and the response [Figs.~\ref{1b_exp_phas1}(b)i and 
\ref{chaos_ls}(b)i] systems for the coupling strength $\varepsilon=0.3$ 
indicate that the time-delay systems evolve independently.
The sets that are localized on the chaotic attractor of both the drive [Figs.~\ref{1b_exp_phas1}(c)i and 
\ref{chaos_ls}(c)i] and the response [Figs.~\ref{1b_exp_phas1}(d)i and 
\ref{chaos_ls}(d)i] systems for the coupling strength $\varepsilon=0.9$ 
correspond to a perfect locking of the phases of both systems
as confirmed by the zero phase difference plotted in  Fig.~\ref{phase_dif}.

%
% Experimental characterization of CPS using the framework of
%localized sets in the hyperchaotic regime for the delay time $\tau=6.0$. 
%(a) and (b) Distributed sets in the drive and the response
%systems, respectively, indicate the asynchronous state for $\varepsilon=3$.
% (c) and (d) Localized sets in the drive and the response
%systems, respectively, indicating the existence of CPS for $\varepsilon=7$.} 
%\label{2b_exp_phas1}
%\end{figure}
%
Next, we confirm the synchronization transition 
using the largest Lyapunov exponents
of the coupled time-delay systems and the $C_{CPR}$~\cite{nmmcr2007}. 
The four largest Lyapunov exponents of (\ref{dd_eq})
are depicted in Fig.~\ref{chaos_ps_char}(a) as a function of
$\varepsilon\in(0,1)$. $(1)$ The zero Lyapunov exponent of the response
system already becomes negative for lower values of $\varepsilon$ and the
positive Lyapunov exponents become gradually negative
for $\varepsilon>0.62$ indicating the existence of CPS.
This is a strong indication of some degree of correlation in the
amplitudes, as transition of positive Lyapunov exponents to negative values correspond
to the stabilization of transverse instabilities of the response attractor,
of both the systems even before the onset of CPS and such a 
negative transition of positive Lyapunov exponents at the onset of CPS
is a typical characteristic of time-delay systems~\cite{dvs2006}.  Similar 
transitions have also been reported in non-phase-coherent attractors of 
low-dimensional systems~\cite{aspmgr2001,sbjk2002,dvs2006}. 
$(2)$ The definition of
$C_{CPR}=\langle \bar{P_1}(t)\bar{P_2}(t)\rangle/\sigma_1\sigma_2$, where 
$\bar{P}_{1,2}$ means that the mean value has been subtracted and
$\sigma_{1,2}$ are the standard deviations of $P_1(t)$ and $P_2(t)$
respectively, $\langle\cdot\rangle$ is the time average and $P(t)$ is a
generalized autocorrelation function based on recurrence properties~\cite{nmmcr2007}.  
If both the systems are in CPS, the probability of recurrence is
maximal at the same time $t$ and $C_{CPR} \approx 1$. If they are not in CPS,
the maxima do not occur simultaneously and hence one can expect a drift in both
the probability of recurrences resulting in low values of  $C_{CPR}$. The low values of
$C_{CPR}$ [Fig.~\ref{chaos_ps_char}(b)] in the range $\varepsilon\in(0,0.62)$ 
indicates that both coupled systems are not in CPS and for $\varepsilon>0.62$
the values of $C_{CPR} \approx 1$ confirming the existence of high quality CPS. 

It is important to note that real time estimation of either of these 
measures is practically not possible. This is because of experimental data 
acquisition with high precision, as a function of all system parameters, 
impose severe limitations on handling huge data set, sampling intervals, 
effect of noise, etc., and even then one has to rely on data analysis tools
for the estimation of both Lyapunov exponents and $C_{CPR}$, which are essentially
numerical analysis.  
Therefore, for the present study, further 
characterizations of CPS using Lyapunov exponents and $C_{CPR}$ are suitably supplemented by 
numerical simulations.
\begin{figure}[!ht]
%\centering\includegraphics[width=0.8\linewidth]{fig1}
\centering\includegraphics[width=0.75\linewidth]{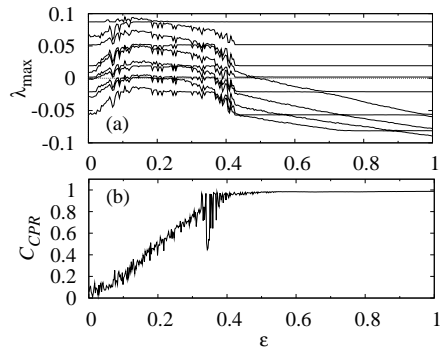}
\caption{Confirmation of CPS in the  hyperchaotic regime using
(a)Ten largest Lyapunov exponents, and (b) Correlation of probability of
recurrence ($C_{CPR}$).} \label{hychaos_ps_char}
\end{figure}

Now, we demonstrate the existence of CPS in a hyperchaotic regime for the delay
time $\tau=6.0$.  For rather samll $\varepsilon$, the sets spread over the 
entire hyperchaotic attractors of the
drive and the response systems. The experimental results are shown in 
Figs.~\ref{1b_exp_phas1}(a)ii and \ref{1b_exp_phas1}(b)ii
and numerical results are given in Figs.~\ref{chaos_ls}(a)ii and \ref{chaos_ls}(b)ii, respectively,
for $\varepsilon=0.3$, which confirm that both systems
evolve independently. On the other hand, for $\varepsilon=0.9$, the observed sets  that are localized on the
hyperchaotic attractors of the drive and the response systems as shown experimentally 
in Figs.~\ref{1b_exp_phas1}(c)ii and \ref{1b_exp_phas1}(d)ii
 and  numerically in Figs.~\ref{chaos_ls}(c)ii and \ref{chaos_ls}(d)ii, respectively,
indeed confirm the existence of CPS in the hyperchaotic regime.
\begin{figure}[!ht]
%\centering\includegraphics[width=0.8\linewidth]{fig1}
\centering\includegraphics[width=0.7\linewidth]{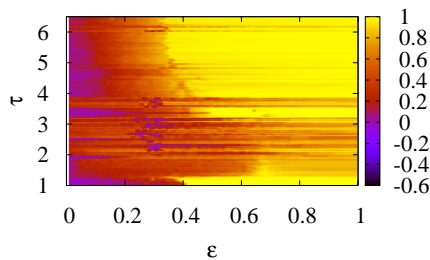}
\caption{(color online) CPS regime in the parameter space $(\varepsilon,\tau)$
characterized using the values of the index $C_{CPR}$.} \label{tp_spec_cpr}
\end{figure}

The largest ten Lyapunov exponents of the coupled time-delay systems for
the delay time $\tau=6.0$ are shown in Fig.~\ref{hychaos_ps_char}(a) in the
range of $\varepsilon\in(0,1)$. The four positive Lyapunov exponents of the
drive system continue to remain positive in the entire range of $\varepsilon$. The three
least positive Lyapunov exponents of the response system become gradually negative
for  $\varepsilon>0.4$ and the largest positive Lyapunov exponent becomes negative
for $\varepsilon>0.5$, at which $C_{CPR}$ [Fig.~\ref{hychaos_ps_char}(b)] also reaches 
the value of unity, indicating the existence of high quality CPS in the
hyperchaotic regime.
Further, we have scanned the $(\varepsilon,\tau)$ parameter space by
calculating the value of $C_{CPR}$ to demarcate the regimes of CPS as 
depicted in Fig.~\ref{tp_spec_cpr}. As discussed above, the coupled systems
are in CPS when the value of $C_{CPR}$$\approx 1$ and it is evident from this figure
that CPS occurs in a wide range of $\tau$.

To summarize, we have demonstrated the notion of CPS in a unidirectionally coupled time-delay
electronic circuit with threshold nonlinearity in both chaotic and hyperchaotic
regimes. The existence of CPS is observed experimentally from snapshots of the time evolution
of both the coupled systems and is confirmed with the framework of localized sets.
Further we have corroborated the synchronization transition numerically using the phase differences,
the concept of localized sets, changes in the largest Lyapunov exponents and  from the 
values of $C_{CPR}$ of the coupled time-delay systems, which agree well with the experimental
observations. We strongly believe that our 
results especially with the framework of localized sets
will lead to the identification of
CPS in other physical systems with delay and to their potential applications.

%\acknowledgments
D. V. S has been supported by the Alexander von Humboldt Foundation. 
The work of K. S. and M. L. has been supported by the Department of Science 
and Technology (DST), Government of India sponsored IRHPA research project, 
INSA senior scientist program, and DST Ramanna program of M. L.  
J. K. acknowledges the support from  EU  under project No. 240763 PHOCUS(FP7-ICT-2009-C).

%\bibliography{sync,chaos}

\begin{thebibliography}{33}
\expandafter\ifx\csname natexlab\endcsname\relax\def\natexlab#1{#1}\fi
\expandafter\ifx\csname bibnamefont\endcsname\relax
  \def\bibnamefont#1{#1}\fi
\expandafter\ifx\csname bibfnamefont\endcsname\relax
  \def\bibfnamefont#1{#1}\fi
\expandafter\ifx\csname citenamefont\endcsname\relax
  \def\citenamefont#1{#1}\fi
\expandafter\ifx\csname url\endcsname\relax
  \def\url#1{\texttt{#1}}\fi
\expandafter\ifx\csname urlprefix\endcsname\relax\def\urlprefix{URL }\fi
\providecommand{\bibinfo}[2]{#2}
\providecommand{\eprint}[2][]{\url{#2}}

\bibitem[{\citenamefont{Pikovsky et~al.}(2001)}]{aspmgr2001}
\bibinfo{author}{\bibfnamefont{A.~S.} \bibnamefont{Pikovsky}},
  \bibinfo{author}{\bibfnamefont{M.~G.} \bibnamefont{Rosenblum}},
  \bibnamefont{and} \bibinfo{author}{\bibfnamefont{J.}~\bibnamefont{Kurths}},
  \emph{\bibinfo{title}{Synchronization - A Unified Approach to Nonlinear
  Science}} (\bibinfo{publisher}{Cambridge University Press},
  \bibinfo{address}{Cambridge}, \bibinfo{year}{2001}).

\bibitem[{\citenamefont{Boccaletti et~al.}(2002)}]{sbjk2002}
\bibinfo{author}{\bibfnamefont{S.}~\bibnamefont{Boccaletti}},
  \bibinfo{author}{\bibfnamefont{J.}~\bibnamefont{Kurths}},
  \bibinfo{author}{\bibfnamefont{G.}~\bibnamefont{Osipov}},
  \bibinfo{author}{\bibfnamefont{D.~L.} \bibnamefont{Valladares}},
  \bibnamefont{and} \bibinfo{author}{\bibfnamefont{C.~S.} \bibnamefont{Zhou}},
  \bibinfo{journal}{Phys. Rep.} \textbf{\bibinfo{volume}{366}},
  \bibinfo{pages}{1} (\bibinfo{year}{2002}).

\bibitem[{\citenamefont{Scannel et~al.}(1999}]{scannell1999}
\bibinfo{author}{\bibfnamefont{J.~W.} \bibnamefont{Scannell et.~al.}},
  \bibinfo{journal}{Cereb. Cortex.} \textbf{\bibinfo{volume}{9}},
  \bibinfo{pages}{277} (\bibinfo{year}{1999}).
%
\bibitem[{\citenamefont{Varela et~al.}(1998)}]{csmgr1998}
\bibinfo{author}{\bibfnamefont{C.} \bibnamefont{Sch\"afer}},
  \bibinfo{author}{\bibfnamefont{M. G.}~\bibnamefont{Rosenblum}},
  \bibinfo{author}{\bibfnamefont{J.}~\bibnamefont{Kurths}},
  \bibnamefont{and} \bibinfo{author}{\bibfnamefont{H. H.}~\bibnamefont{Abel}},
  \bibinfo{journal}{Nature} \textbf{\bibinfo{volume}{392}},
  \bibinfo{pages}{239} (\bibinfo{year}{1998}).
%
\bibitem[{\citenamefont{Varela et~al.}(1999)}]{bbah1999}
\bibinfo{author}{\bibfnamefont{B.} \bibnamefont{Blasius}},
  \bibinfo{author}{\bibfnamefont{A.}~\bibnamefont{Huppert}},
  \bibnamefont{and} \bibinfo{author}{\bibfnamefont{L.}~\bibnamefont{Stone}},
  \bibinfo{journal}{Nature} \textbf{\bibinfo{volume}{399}},
  \bibinfo{pages}{354} (\bibinfo{year}{1999}).
%
\bibitem[{\citenamefont{Grenfell et~al.}(2001}]{grenfell2001}
\bibinfo{author}{\bibfnamefont{B.~T.} \bibnamefont{Grenfell et.~al.}},
  \bibinfo{journal}{Nature (London)} \textbf{\bibinfo{volume}{414}},
  \bibinfo{pages}{716} (\bibinfo{year}{2001}).

\bibitem[{\citenamefont{RosaJr. et~al.}(2003)\citenamefont{Rosa, Ticos,
  Pardo, Walkenstein, Monti, and Kurths}}]{ercmt2003}
\bibinfo{author}{\bibfnamefont{E.}~\bibnamefont{Rosa et.~al.}},
  \bibinfo{journal}{Phys. Rev. E} \textbf{\bibinfo{volume}{68}},
  \bibinfo{pages}{025202(R)} (\bibinfo{year}{2003});
\bibinfo{author}{\bibfnamefont{M.~S.} \bibnamefont{Baptista et.~al.}},
  \bibinfo{journal}{Phys. Rev. E} \textbf{\bibinfo{volume}{67}},
  \bibinfo{pages}{056212} (\bibinfo{year}{2003}).

\bibitem[{\citenamefont{Volodehenko et~al.}(2001)\citenamefont{Volodehenko,
  Ivanov, Gong, Choi, Park, and Kim}}]{kvvvni2001}
\bibinfo{author}{\bibfnamefont{K.~V.} \bibnamefont{Volodehenko et.~al.}},
  \bibinfo{journal}{Opt. Lett.} \textbf{\bibinfo{volume}{26}},
  \bibinfo{pages}{1406} (\bibinfo{year}{2001});
\bibinfo{author}{\bibfnamefont{D.~J.} \bibnamefont{DeShazer et.~al.}},
  \bibinfo{journal}{Phys. Rev. Lett.} \textbf{\bibinfo{volume}{87}},
  \bibinfo{pages}{044101} (\bibinfo{year}{2001}).

\bibitem[{\citenamefont{Maza et~al.}(2000)\citenamefont{Maza, Vallone, Mancini,
  and Boccaletti}}]{dmav2000}
\bibinfo{author}{\bibfnamefont{D.}~\bibnamefont{Maza}},
  \bibinfo{author}{\bibfnamefont{A.}~\bibnamefont{Vallone}},
  \bibinfo{author}{\bibfnamefont{H.}~\bibnamefont{Mancini}}, \bibnamefont{and}
  \bibinfo{author}{\bibfnamefont{S.}~\bibnamefont{Boccaletti}},
  \bibinfo{journal}{Phys. Rev. Lett.} \textbf{\bibinfo{volume}{85}},
  \bibinfo{pages}{5567} (\bibinfo{year}{2000}).

\bibitem[{\citenamefont{Tass et~al.}(1998)\citenamefont{Tass, Rosenblum, Weule,
  Kurths, Pikovsky, Volkmann, Schnitzler, and Freund}}]{ptmgr1998}
\bibinfo{author}{\bibfnamefont{P.}~\bibnamefont{Tass et.~al.}},
\bibinfo{journal}{Phys. Rev. Lett.}
  \textbf{\bibinfo{volume}{81}}, \bibinfo{pages}{3291} (\bibinfo{year}{1998});
%\bibitem[{\citenamefont{Elson et~al.}(1998)\citenamefont{Elson, Selverston,
%  Huerta, Rulkov, Rabinovich, and Abarbanel}}]{rceals1998}
\bibinfo{author}{\bibfnamefont{R.~C.} \bibnamefont{Elson et.~al.}},
\bibinfo{journal}{Phys. Rev. Lett.}
  \textbf{\bibinfo{volume}{81}}, \bibinfo{pages}{5692} (\bibinfo{year}{1998}).

\bibitem[{\citenamefont{Maraun and Kurths}(2005)}]{dmjk2005}
\bibinfo{author}{\bibfnamefont{D.}~\bibnamefont{Maraun}} \bibnamefont{and}
  \bibinfo{author}{\bibfnamefont{J.}~\bibnamefont{Kurths}},
  \bibinfo{journal}{Geophys. Res. Lett.} \textbf{\bibinfo{volume}{32}},
  \bibinfo{pages}{L15709} (\bibinfo{year}{2005}).

\bibitem[{\citenamefont{Vicent et~al.}(1998)\citenamefont{Vicent,
  Tang, Mulet, Mirasso,  and Liu}}]{rvst2004}
\bibinfo{author}{\bibfnamefont{P.}~\bibnamefont{Vicent et.~al.}},
\bibinfo{journal}{Phys. Rev. E}
  \textbf{\bibinfo{volume}{70}}, \bibinfo{pages}{046216} (\bibinfo{year}{2004}).

\bibitem[{\citenamefont{Senthil2 et~al.}(2006)}]{dvs2006}
\bibinfo{author}{\bibfnamefont{D.~V.} \bibnamefont{Senthilkumar}},
  \bibinfo{author}{\bibfnamefont{M.}~\bibnamefont{Lakshmanan}},
  \bibnamefont{and} \bibinfo{author}{\bibfnamefont{J.}~\bibnamefont{Kurths}},
  \bibinfo{journal}{Phys. Rev. E} \textbf{\bibinfo{volume}{74}},
  \bibinfo{pages}{035205(R)} (\bibinfo{year}{2006});
  \bibinfo{journal}{Chaos} \textbf{\bibinfo{volume}{18}},
  \bibinfo{pages}{023118} (\bibinfo{year}{2008});  
  \bibinfo{journal}{Eur. Phys. J. Special Topics} \textbf{\bibinfo{volume}{164}},
  \bibinfo{pages}{35} (\bibinfo{year}{2008}).

\bibitem[{\citenamefont{Suresh et~al.}(2010)}]{rsdvs2010}
\bibinfo{author}{\bibfnamefont{R.} \bibnamefont{Suresh}},
\bibinfo{author}{\bibfnamefont{D.~V.} \bibnamefont{Senthilkumar}},
  \bibinfo{author}{\bibfnamefont{M.}~\bibnamefont{Lakshmanan}},
  \bibnamefont{and} \bibinfo{author}{\bibfnamefont{J.}~\bibnamefont{Kurths}},
  \bibinfo{journal}{Phys. Rev. E} \textbf{\bibinfo{volume}{82}},  
\bibinfo{pages}{016215} (\bibinfo{year}{2010}).


\bibitem[{\citenamefont{Pereira et~al.}(2007)}]{tpmsb2007}
\bibinfo{author}{\bibfnamefont{T.} \bibnamefont{Pereira}}, 
  \bibinfo{author}{\bibfnamefont{M.~S.} \bibnamefont{Baptista}}, 
\bibnamefont{and} \bibinfo{author}{\bibfnamefont{J.} \bibnamefont{Kurths}}, 
  \bibinfo{journal}{Phys. Rev. E} \textbf{\bibinfo{volume}{75}},
  \bibinfo{pages}{026216} (\bibinfo{year}{2007}).

\bibitem[{\citenamefont{Marwan et~al.}(2007)}]{nmmcr2007}
\bibinfo{author}{\bibfnamefont{N.} \bibnamefont{Marwan}},
\bibinfo{author}{\bibfnamefont{M. C.} \bibnamefont{Romano}},
\bibinfo{author}{\bibfnamefont{M.} \bibnamefont{Thiel}},
  \bibnamefont{and}
  \bibinfo{author}{\bibfnamefont{J.} \bibnamefont{Kurths}},
  \bibinfo{journal}{Phys. Rep.} \textbf{\bibinfo{volume}{438}},
  \bibinfo{pages}{237} (\bibinfo{year}{2007}). 

\bibitem[{\citenamefont{Srinivasan et~al.}(2007)\citenamefont{Srinivasan, Raja Mohamed, Murali, Lakshmanan and Sudeshna Sinha}}]{srini2010}
\bibinfo{author}{\bibfnamefont{K.}~\bibnamefont{Srinivasan et.~al.}}, 
  \bibinfo{journal}{Int. J. Bifurcation and Chaos} \textbf{\bibinfo{volume}{21}},
(\bibinfo{year}{2011})(To appear); arXiv:1008.4011.


\bibitem[{\citenamefont{Srinivasan et~al.}(2007)\citenamefont{Srinivasan, Murali, Lakshmanan and Kurths}}]{ksdvs2010}
\bibinfo{author}{\bibfnamefont{K.}~\bibnamefont{Srinivasan}},
\bibinfo{author}{\bibfnamefont{D.~V.} \bibnamefont{Senthilkumar}},
  \bibinfo{author}{\bibfnamefont{K.}~\bibnamefont{Murali}},
  \bibinfo{author}{\bibfnamefont{M.}~\bibnamefont{Lakshmanan}},
  \bibnamefont{and} \bibinfo{author}{\bibfnamefont{J.}~\bibnamefont{Kurths}},
  \bibinfo{journal}{(submitted)}; arXiv:1008.3300.
%
%
\end{thebibliography}

\end{document}